\documentclass{appolb}
\usepackage{epsfig}
\def\be{\begin{equation}}
\def\ee{\end{equation}}
\def\bea{\begin{eqnarray}}
\def\eea{\end{eqnarray}}

\newcommand{\scs}{\scriptscriptstyle}
\newcommand{\f}{\frac}

\newcommand{\gsim}{\;\rlap{\lower 4.5 pt \hbox{$\mathchar \sim$}} \raise 2pt \hbox {$>$}\;}
\newcommand{\lsim}{\;\rlap{\lower 4.5 pt \hbox{$\mathchar \sim$}} \raise 2pt \hbox {$<$}\;}
\begin{document}
\title{ $\bar B \to X_s \gamma$ -- Current Status
\thanks{Presented at the XXXIII International Conference on Theoretical Physics: 
``Matter to the Deepest'', Ustro\'n, Poland, September 11-16, 2009.}}
\author{Miko{\l}aj Misiak
\address{Institute of Theoretical Physics, University of Warsaw,\\ 
ul. Ho\.za 69, 00-681 Warsaw, Poland.}}
\maketitle
\begin{abstract}
Our current knowledge of ${\cal B}(\bar B \to X_s \gamma)$ is briefly
summarized, with particular attention to uncertain non-perturbative effects. 
\end{abstract}
\PACS{12.38.Bx, 13.20.He}

\section{Introduction \label{sec:intro}}

Weak radiative $\bar B$-meson decays ($\bar B = \bar B^0$ or $B^-$) are
generated by the Flavour-Changing Neutral Current (FCNC) $b \to s \gamma$
transition that arises at one loop only. The diagrams contain sums over all the
up-type quark flavours that are highly non-degenerate in mass.  The relevant
Cabibbo-Kobayashi-Maskawa (CKM) factor $|V_{ts}^* V_{tb}|$ is very close to
$|V_{cb}|$ that occurs in the leading $b$-quark decays. In effect, the usual
Glashow-Iliopoulos-Maiani suppression mechanism for FCNC processes is not at
work. Numerically, the inclusive branching ratio~ ${\cal B}_{s\gamma} \equiv
{\cal B}(\bar B \to X_s \gamma) \simeq 0.14\; \f{\alpha_{\rm em}}{\pi}$~ in
the Standard Model (SM).

Contributions of potentially the same size arise in extensions of the SM. \linebreak  
In supersymmetric models, for instance, 
$\Delta {\cal B}_{s\gamma}^{\rm SUSY}/{\cal B}_{s\gamma}^{\rm SM}$
scales roughly like 
$\left[ 100\;{\rm GeV}/\left(\mbox{superpartner masses}\right)\right]^2$.
Consequently, comparing measurements of ${\cal B}_{s\gamma}$ with theory
predictions at a few percent level gives us strong constraints on new physics,
irrespectively of whether any deviations from the SM are observed.

Accurate determination of ${\cal B}_{s \gamma}$ is challenging on both the
experimental and theoretical sides. At the $B$-factories, one of the main
difficulties is precise subtraction of the so-called continuum background,
i.e. hard photons that originate from non-$B\bar B$ processes. Both the
continuum and the $\bar B \to X_c \gamma$ backgrounds become more severe
towards lower photon energies. It is clearly reflected by the behaviour of
errors in the most recent BELLE measurement~\cite{Limosani:2009qg}
of the (isospin- and CP-averaged) branching ratio
\be \label{belle}
{\cal B}_{s \gamma}(E_\gamma > E_0) = \left\{ \begin{array}{ll}
\left( 3.02 \pm 0.10_{\rm stat} \pm 0.11_{\rm syst} \right)\times 10^{-4},
& \mbox{~for~} E_0 = 2.0\;{\rm GeV},\\[1mm]
\left( 3.21 \pm 0.11_{\rm stat} \pm 0.16_{\rm syst} \right)\times 10^{-4},
& \mbox{~for~} E_0 = 1.9\;{\rm GeV},\\[1mm]
\left( 3.36 \pm 0.13_{\rm stat} \pm 0.25_{\rm syst} \right)\times 10^{-4},
& \mbox{~for~} E_0 = 1.8\;{\rm GeV},\\[1mm]
\left( 3.45 \pm 0.15_{\rm stat} \pm 0.40_{\rm syst} \right)\times 10^{-4},
& \mbox{~for~} E_0 = 1.7\;{\rm GeV}.
\end{array} \right.
\ee
On the other hand, the theory prediction is based on an approximate equality
of the hadronic and partonic decay widths
\be \label{main}                                
\Gamma(\bar B \to X_s \gamma)_{{}_{E_\gamma > E_0}}
~\simeq~ \Gamma(b \to X_s^{\rm parton} \gamma)_{{}_{E_\gamma > E_0}}
\ee
that breaks down when $E_0$ is too close to the endpoint 
$E_{\rm max} = \f{m_B^2 - m_K^{*\;2}}{2 m_B} \simeq 2.56\;$GeV,
i.e. when~ $E_{\rm max} - E_0 \sim \Lambda \equiv\Lambda_{\scs\rm QCD}$.
It has become customary to use $E_0 = 1.6\;{\rm GeV} \simeq \f{m_b}{3}$
for comparing theory with experiment. The SM prediction for this value
of $E_0$ reads~\cite{Misiak:2006zs}
\be \label{sm}
{\cal B}_{s \gamma}(E_\gamma > 1.6\;{\rm GeV}) = \left( 3.15 \pm 0.23 \right)\times 10^{-4}. 
\ee
It includes the ${\cal O}(\alpha_s^2)$ QCD corrections and the leading
electroweak ones. Both the known and unknown non-perturbative effects have
been taken into account in estimating the central value and the uncertainty.

The currently available experimental world averages read
\be \label{aver}
{\cal B}_{s \gamma}(E_\gamma > 1.6\;{\rm GeV}) = \left\{ \begin{array}{l}
\left( 3.52 \pm 0.23_{\rm exp} \pm 0.09_{\rm model} \right)\times 10^{-4}~~~~\cite{Barberio:2008fa},
\\[1mm]
\left( 3.50 \pm 0.14_{\rm exp} \pm 0.10_{\rm model} \right)\times 10^{-4}~~~~\cite{Artuso:2009jw}.
\end{array} \right.
\ee
They have been obtained by extrapolation from measurements at higher $E_0$
using models of the photon energy spectrum and fitting their parameters to
data. The BELLE result in Eq.~(\ref{belle}) is more recent than the above
averages. Its preliminary version~\cite{Abe:2008sxa} has been used in
Eq.~(\ref{aver}) instead, together with older measurements of CLEO, BELLE and
BABAR~\cite{Chen:2001fj,Aubert:2005cu}.\footnote{
  The average in Ref.~\cite{Barberio:2008fa} has a larger error because it
  includes results at $E_0 \geq 1.8\;$GeV from the older
  measurements~\cite{Chen:2001fj,Aubert:2005cu} only, ignoring the more
  precise ones from Ref.~\cite{Abe:2008sxa}.}
%
%
%
The SM prediction (\ref{sm}) and the averages (\ref{aver}) are consistent
at the $1.2\sigma$ level.

Theoretical analyses of~ $\bar B \to X_s \gamma$~ employ the formalism of an
effective theory that arises after decoupling of the $W$-boson and all the
heavier particles. The relevant flavour-changing interactions are given by
dimension-five and -six local operators\footnote{
  The specific matrices $\Gamma_i$ and $\Gamma'_i$ can be found in
  Ref.~\cite{Chetyrkin:1996vx}. Additional operators may arise beyond SM.}
\be \label{ops} \begin{array}{rclrcl}
O_{1,2} &=& (\bar{s} \Gamma_i c)(\bar{c} \Gamma'_i b), \hspace{1cm} &
O_{3,4,5,6} &=&  (\bar{s} \Gamma_i b) {\textstyle \sum_q} (\bar{q} \Gamma'_i q),\\[2mm]
O_7 &=& \f{e m_b}{16 \pi^2}\, \bar{s}_L \sigma^{\mu \nu} b_R F_{\mu \nu}, &
O_8 &=& \f{g m_b}{16 \pi^2}\, \bar{s}_L \sigma^{\mu \nu} T^a b_R G^a_{\mu\nu}. 
\end{array} \ee 
One begins with perturbatively calculating their Wilson coefficients $C_i$ at
the renormalization scale $\mu_0 \sim (M_W, m_t)$. Next, renormalization group
is used for the evolution of $C_i$ down to the scale $\mu_b \sim m_b/2$.
These two steps were finalized a few years ago including ${\cal
  O}(\alpha_s^2)$ effects in the SM~\cite{Bobeth:1999mk}. The last step
amounts to evaluating the inclusive decay width $\Gamma(\bar B \to X_s
\gamma)$ that is generated by $Q_i$ at the scale $\mu_b$. Non-perturbative
effects show up at this last stage only.

\begin{figure}[b]
\hspace*{-5mm}
\psfig{figure=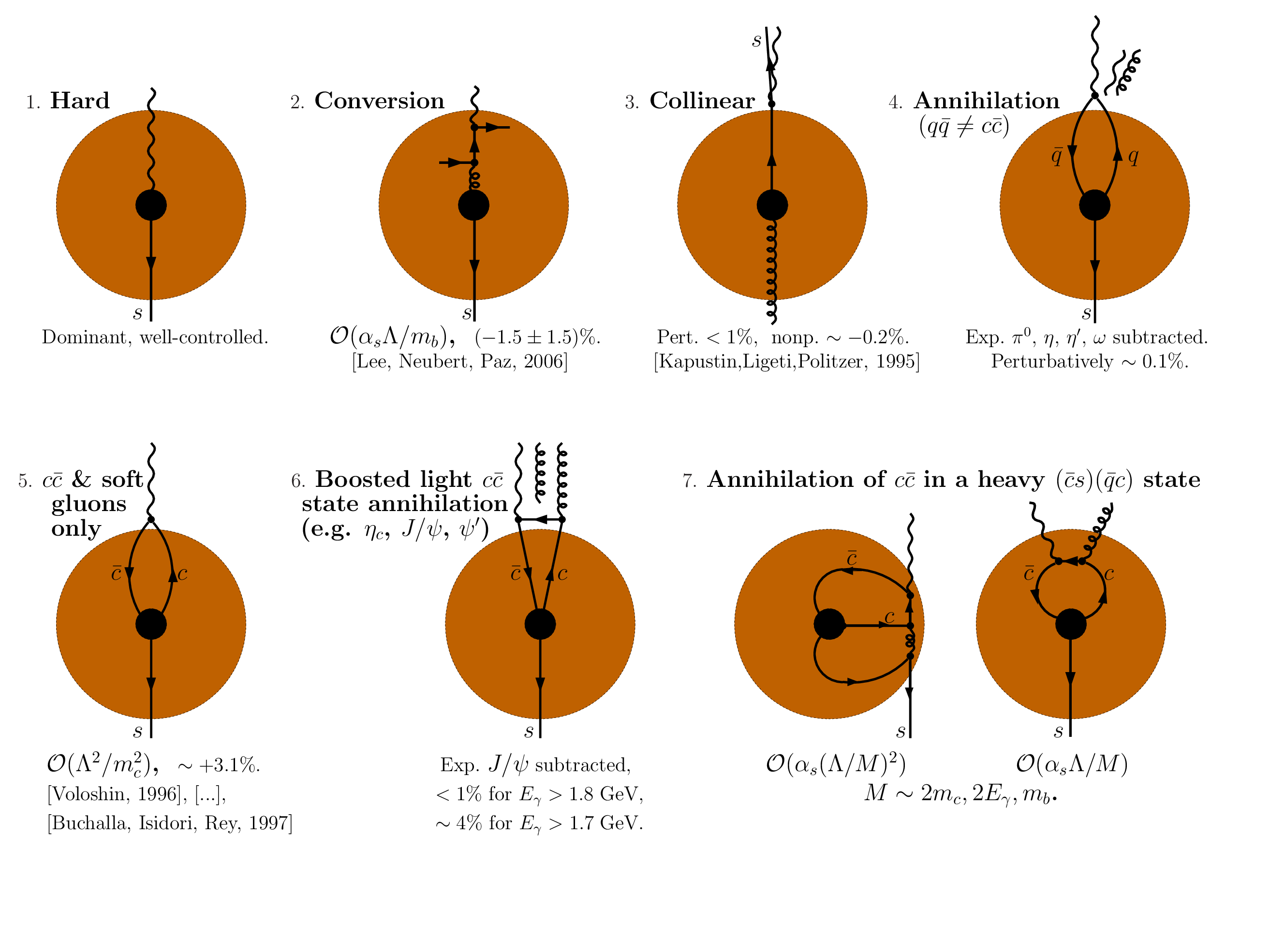,width=14cm}\\[-12mm]
\caption{Energetic photon production ($E_\gamma \gsim \f{m_b}{3}$) in
  charmless decays of the $\bar B$ meson. The first and second rows describe
  situations without and with long-distance charm loops, respectively (see the
  text).}
\label{fig:rings}
\end{figure}

The final states in~ $\bar B \to X_s \gamma$~ are required to be charmless,
i.e. to contain no charmed ($C \neq 0$) hadrons. Various ways of energetic
photon production ($E_\gamma \gsim \f{m_b}{3}$) in $\bar B$-meson decays to
such states are displayed in Fig.~\ref{fig:rings}.  The first and second rows
describe situations without and with long-distance charm loops,
respectively. The meson in its rest frame is shown as a brown (grey) ball of
diameter $\sim \Lambda^{-1}$, while the heavy $b$-quark is localized in the
center at distances of order $m_b^{-1}$ (black blob). Decay products are
depicted diagrammatically, and their hadronization is implicitly assumed.

\section{Contributions without long-distance charm loops}
\subsection{Hard \label{subsec:hard}}

In the first diagram in Fig.~\ref{fig:rings}, the photon is emitted directly
from the hard process of the $b$-quark decay. This means that the photon
emission vertex and the $b$-quark annihilation vertex either coincide in
position space (as in $Q_7$ in Eq.~(\ref{ops})) or get connected by a (chain
of) high-virtuality parton propagator(s) (virtuality $\sim m_b^2$ or
larger). In the latter case, Operator Product Expansion (OPE) can be performed
at the amplitude level to make the two vertices coincide. Next, in analogy to
the semileptonic $B$-meson decays~\cite{Chay:1990da}, the optical theorem and
another OPE are applied to show that
\be \label{main.hard}                                
\Gamma(\bar B \to X_s \gamma)^{\rm hard}_{{}_{E_\gamma > E_0}}
~=~ \Gamma(b \to X_s^{\rm parton} \gamma)^{\rm hard}_{{}_{E_\gamma > E_0}} ~+~ 
{\cal O}\left(\f{\Lambda^2}{m_b^2}\right),
\ee
so long as ~$(m_b - 2 E_0) \sim m_b$.~ The 
${\cal O}\left(\Lambda^2/m_b^2\right)$
corrections are expressed~\cite{Bigi:1992ne} in terms of local operator matrix
elements between the $\bar B$-meson states at rest. These matrix elements can be
extracted from $B$--$B^*$ mass splitting and/or from properties of the
semileptonic decay spectra (see e.g. Ref.~\cite{Schwanda:2008kw}).
Numerically, the
${\cal O}\left(\Lambda^2/m_b^2\right)$
corrections amount to around $-3\%$ of the decay rate.

Other-than-hard contributions to $\Gamma(b \to X_s^{\rm parton} \gamma)$ arise
only inside perturbative corrections of order ${\cal O}(\alpha_s)$ and higher.
In the actual calculations of these corrections, all the momentum regions are
included, i.e. hard contributions are not singled out. Consequently, the
remaining non-perturbative effects to be discussed below should be understood
as containing a subtraction of the corresponding perturbative terms (if
present).

\subsection{Conversion}

In the second diagram of Fig.~\ref{fig:rings}, the $b$-quark decays in a hard
way to quarks and gluons only. Next, one of the decay products scatters
radiatively with remnants of the $\bar B$-meson in a non-soft manner,
i.e. with momentum transfer much larger than $\Lambda$. Such a situation
should be distinguished from collinear photon emission in jet fragmentation
where only soft interactions are sufficient (see the next subsection). Here,
contributions to the decay rate are suppressed with respect to the leading
hard~ $b \to s\gamma$~ one by $\alpha_s$ (due to non-soft scattering) and by
$\Lambda/m_b$ (due to dilution of the target). The analysis of
Ref.~\cite{Lee:2006wn} confirms that no other suppression factors occur.

The considered amplitudes should efficiently interfere with the leading hard~
$b \to s\gamma$~ ones to make the effect relevant, i.e. both the photon and
the $s$-quark should move roughly back-to-back with energies close to $m_b/2$.
For this reason, we can restrict ourselves to the gluon-to-photon
case\footnote{
 This very case has been displayed in the second diagram of Fig.~\ref{fig:rings}.
 The analogous quark-to-photon transition could hardly give enough interference.}
that resembles Compton scattering, and think of the scattered quark as soft on
both external lines. This quark may either be a valence quark or any of the
sea quarks. Thus, we deal with a process that may be viewed as gluon-to-photon
conversion in the QCD medium.

In Ref.~\cite{Lee:2006wn}, the Vacuum Insertion Approximation (VIA) has been
used to find that conversion gives a correction
$\Delta {\cal B}_{s\gamma}/{\cal B}_{s\gamma} \in [-3, -0.3]\%$ 
to the isospin averaged branching ratio
\be
{\cal B}_{s\gamma} ~=~ 
\f{\Gamma(\bar B^0 \to X_s \gamma) + \Gamma(B^- \to X_s \gamma)}{
   \Gamma_{\rm tot}(\bar B^0) + \Gamma_{\rm tot}(B^\pm)} ~\equiv~ 
\f{\Gamma^0_{s\gamma} + \Gamma^-_{s\gamma}}{
   \Gamma^0_{\rm tot} + \Gamma^-_{\rm tot}},
\ee
and generates a sizeable isospin asymmetry
$\Delta_{0-} = (\Gamma^0_{s\gamma} - \Gamma^-_{s\gamma})/(\Gamma^0_{s\gamma} + \Gamma^-_{s\gamma})$.
However, assuming a larger uncertainty of~ $\sim\!\!5\%$ in 
$\Delta {\cal B}_{s\gamma}/{\cal B}_{s\gamma}$ 
has been recommended in view of the fact that VIA is a very rough approximation.

It is interesting to note that instead of the VIA one may consider 
the $SU(3)_{\rm flavor}$ limit that amounts to neglecting differences
between light quark masses as well as electromagnetic corrections to the meson
wave-functions. In this limit,~
$\Gamma^0_{s\gamma} \simeq \Gamma_{\rm (no~conversion)} + Q_d \Delta \Gamma_1 + (Q_u + Q_s) \Delta \Gamma_2$~
and~
$\Gamma^-_{s\gamma} \simeq \Gamma_{\rm (no~conversion)} + Q_u \Delta \Gamma_1 + (Q_d + Q_s) \Delta \Gamma_2$,~
which follows from the fact that interference of the gluon-to-photon
conversion with the leading hard amplitude is linear in the quark charges.
Next, using~ $Q_u+Q_d+Q_s=0$,~ one immediately finds
\be
\Delta {\cal B}_{s\gamma}/{\cal B}_{s\gamma} 
\simeq~ \f{Q_d+Q_u}{Q_d-Q_u}~ \Delta_{0-}
= -\f{1}{3} \Delta_{0-} = \left( +0.2 \pm 1.9_{\rm stat}
\pm 0.3_{\rm sys} \pm 0.8_{\rm ident} \right)\%,
\ee
where the BABAR measurement~\cite{Aubert:2005cu} of $\Delta_{0-}$ for
$E_\gamma > 1.9\,$GeV has been used in the last step.  It seems reasonable to
assume that the valence quark effects in
$\Delta {\cal B}_{s\gamma}/{\cal B}_{s\gamma}$ 
dominate over the $SU(3)_{\rm flavor}$-violating ones. Thus, if the more
precise future measurements of $\Delta_{0-}$ remain consistent with zero, our
control over uncertain contributions to ${\cal B}_{s\gamma}$ may significantly
improve.

\subsection{Collinear}

In the third diagram of Fig.~\ref{fig:rings}, a collinear photon is
emitted in the process of hadronization. We require $|\vec{p}\,|_\gamma >
1.6\;{\rm GeV} \sim m_b/3$.  The maximal parton three-momentum in the $b$-quark
decay is $m_b/2$, but a typical one is much lower. Thus, there is rather
little phase-space for a collinear emission.\footnote{
  The phase-space could be further reduced in the future by performing the
  $E_0$-extrapolation solely for the hard contributions. The remaining ones
  could be subtracted from the experimental data at higher $E_0$ in advance.}
Moreover, charmless hadronic $B$-decays are parametrically suppressed either
by $\alpha_s$ (for $Q_8$), or by the small Wilson coefficients ($\lsim 0.07$
for $Q_{3,4,5,6}$), or by the CKM angles (for the $u$-quark analogues of
$Q_{1,2}$).

Perturbatively, $(n > 2)$-body decays that involve operators other than $Q_7$~
are responsible for~ $4\div6\%$~ of~ $\Gamma(b \to X_s^{\rm parton} \gamma)$~
for~ $E_0 \in [1.6,2.0]\;$GeV.\linebreak A small fraction of them involves
collinear configurations. In $b \to sg\gamma$, a collinear logarithm ($\ln
m_b/m_s \simeq \ln 50$) arises at ${\cal O}(\alpha_s)$ only in the 88-term,
where ``$kj$-term'' stands for a product of amplitudes generated by $Q_k$ and
$Q_j$.  For $E_\gamma > 1.6\;$GeV, the decay rate changes by less than 1\% or
0.2\% when this logarithm is respectively set to zero or modified according to
Ref.~\cite{Kapustin:1995fk}. Fragmentation functions have been used in that
paper to determine the collinear ($\ln m_b$)-terms. Such small numerical
effects imply that uncertainties in $B_{s\gamma}$ that are due to
non-perturbative collinear effects can be safely neglected at present.

\subsection{Annihilation of light quarks ($q\bar q \neq c\bar c$)}

The fourth diagram in Fig.~\ref{fig:rings} describes radiative annihilation of
light quark pairs that have been produced in the $\bar B$-meson
decay. Perturbatively, such effects amount to only around $0.1\%$ of the decay
rate. In reality, they are very much enhanced for several lightest $q\bar q$
mesons ($\pi^0$, $\eta$, $\eta'$, $\omega$, $\rho$) because of the limited
number (or lack) of alternative hadronic decay channels. However, photons
originating from radiative decays of $\pi^0$ and $\eta$ are vetoed on the
experimental side. Radiative decays of other light mesons are simulated and
treated as background, too. Thus, it is mandatory to assume that all the
non-perturbatively enhanced radiative $q\bar q$ annihilation processes are
removed from the signal. On the other hand, the perturbative contributions are
so tiny that retaining them in the theory calculations does not hurt.

\section{Contributions with long-distance charm loops}

Radiative annihilation of intermediate $c\bar c$ states requires much more
care, which is signaled by the presence of large perturbative charm loop
contributions. It is sufficient to mention that changing the charm quark mass
from its measured value~ $m_c(m_c) \simeq 1.28\,$GeV \cite{Chetyrkin:2009fv}
to~ $m_c=m_b$~ results in a suppression of $\Gamma(b \to X_s^{\rm parton}
\gamma)$~ by around 35\% (!). 
Thus, it is important to verify what fraction of ${\cal B}_{s\gamma}$
originates from long-distance $c\bar c$ loops, i.e. from those intermediate
$c\bar c$ states that are not properly accounted for in the perturbative
approach.

We do not need to worry about purely hadronic annihilation of intermediate
$c\bar c$ states because such processes become kinematically allowed in~ 
$\bar B \to X_s \gamma$~ only for~ 
$E_\gamma < (m_B^2-(m_{\eta_c}+m_K)^2)/(2 m_B) \simeq 1.5\;$GeV.

\subsection{A loop with soft gluons only \label{subsec:ccsoft}}

Let us first consider those $c\bar c$ contributions that are not suppressed by
$\alpha_s$, i.e. no hard gluons are present. They are represented by the fifth
diagram in Fig.~\ref{fig:rings} where soft-gluon dressing is implicitly
assumed. Bringing the two charm quarks close to their mass shell in the
intermediate state implies that the momentum $p_g$ absorbed by the soft gluon
field in the $c\bar c$ annihilation process must satisfy $(p_g + p_\gamma)
\sim 4 m_c^2$. Thus, for~ $m_c^2 \gg m_b \Lambda$,~ the on-shell-like
configuration is kinematically inaccessible, which implies that the charm
quark loop is a short-distance one in this limit. Its interaction with the
soft gluon field can be expressed in terms of a series of local operators.

In reality, the inequality $m_c^2 > m_b \Lambda$ is barely
satisfied. Consequently, one should worry about all orders of the
OPE. Moreover, as the relevant QCD interactions are soft, one should treat
$\alpha_s$ as a quantity of order unity despite using the language of Feynman
diagrams. As follows from Ref.~\cite{Buchalla:1997ky}, the $c\bar c$ loop with
a single external soft gluon\footnote{
The loop with no external gluons vanishes for the on-shell photon.}
gives a correction to ${\cal B}_{s\gamma}$ that can be written as
\be \label{volcor}
\f{\Lambda^2}{m_c^2} \sum_{n=0}^\infty b_n \left( \f{\Lambda m_b}{m_c^2} \right)^n
\ee
Each $\Lambda^n$ above should be understood as matrix element of a local
operator between $\bar B$-meson states at rest. The numerical coefficients
$b_n$ are found by Taylor-expanding the $c\bar c$ loop diagram in the soft
gluon external momentum. Larger numbers of gluons bring higher powers of
$\Lambda^2/m_c^2$.

The leading $n=0$ term in Eq.~(\ref{volcor}) is calculable because it involves
the same hadronic matrix element as the ${\cal O}(\Lambda^2/m_b^2)$ correction
in Sec.~\ref{subsec:hard}. Matrix elements entering at $n \geq 1$ remain
unknown. However, the coefficients $b_n$ are found to decrease rapidly with
$n$~\cite{Buchalla:1997ky}. Consequently, the $n=0$ term is believed to
provide a good approximation to the whole series. The corresponding correction
to ${\cal B}_{s\gamma}$ amounts to around $+3.1\%$.

\subsection{Boosted light $c\bar c$ state annihilation}

If a hard gluon is attached to the $c\bar c$ loop, we get a suppression by
$\alpha_s$ but need to deal with charm quarks close to their mass
shell. Consequently, techniques from the previous section become inapplicable.
Now the charm quarks can form a light $c\bar c$ meson like $J/\psi$ (see the
sixth diagram in Fig.~\ref{fig:rings}) whose radiative decay modes are
enhanced due to its tiny hadronic decay width. The measured inclusive
branching ratios ${\cal B}(\bar B \to X_s J/\psi) \simeq (1.094 \pm
0.032)\%$~\cite{Amsler:2008zzb} and ${\cal B}(J/\psi \to gg\gamma) \simeq (9.2
\pm 1.0)\%$~\cite{Besson:2008pr} illustrate the relevance of the intermediate
$J/\psi$ state. 

According to the current conventions, photons originating from radiative
$J/\psi$ decays are treated as background in $\bar B \to X_s \gamma$
measurements. The high photon energy cut suppresses this background by orders
of magnitude with respect to what the above-mentioned total branching ratios
might suggest. For the BELLE results quoted in Eq.~(\ref{belle}), the
intermediate $J/\psi$ background is as large as 4\% and 1\% of the signal for
$E_0 = 1.7\;$GeV and $1.8\;$GeV, respectively~\cite{Limosani:2008priv}. It has
been calculated using Monte Carlo simulations based on the measured inclusive
${\cal B}(\bar B \to X_s J/\psi)$ spectra and on the rich
PDG~\cite{Besson:2008pr} collection of $J/\psi$ exclusive radiative decay
modes.

The same prescription should actually be applied to all the $c\bar c$
resonances that lay below the $D\bar D$ production threshold.\footnote{
  New measurements of inclusive radiative $J/\psi$ and $\psi'$ decay spectra
  have recently become available~\cite{Besson:2008pr}.}
The corresponding background subtraction (with respect to that for $J/\psi$)
is expected to be about 6 times smaller for $\psi'$ and many times smaller for
the remaining (much wider) resonances. With the present uncertainties, the
only non-negligible subtraction is the already included $J/\psi$ one.

The last question to be posed in this subsection is whether anything should be
accordingly subtracted from the perturbative calculation of $\Gamma(b \to
X_s^{\rm parton} \gamma)$.  All the $b \to sg \gamma$ diagrams with charm
loops at ${\cal O}(\alpha_s)$ and ${\cal O}(\alpha_s^2\beta_0)$ contribute to
this width at the levels of 3.6\%, 3.3\% and 2.9\% for $E_0=1.6$, $1.7$ and
$1.8\;$GeV, respectively. Only small fractions of these numbers should
correspond to narrow resonances because the diagrams are calculated in fixed
order, without any Coulomb ladder resummation~\cite{Beneke:2009az}. Thus,
including the perturbative contributions with no subtraction at all seems to
be the right choice.

\subsection{Annihilation of $c\bar c$ in a heavy $(\bar c s)(\bar q c)$ state}

The last two diagrams in Fig.~\ref{fig:rings} are supposed to represent
radiative annihilation of $c\bar c$ pairs in heavy $(\bar c s)(\bar q c)$
states, i.e. in all the states lay above the $D\bar D$ production
threshold. Here, we include not only $c\bar c$-meson-like states, but also
rescattering of heavy $C \neq 0$ hadrons, or charm quark annihilation that
takes place before hadronization. Contributions from such intermediate states
are largely accounted for by the perturbative results and/or those already
discussed in Sec.~\ref{subsec:ccsoft}. However, non-perturbative rescattering
effects in particular exclusive modes may be sizeable. For instance, branching
ratios~\cite{Amsler:2008zzb}~
${\cal B}( B^- \to D_{sJ}(2457)^-\;D^{*}(2007)^0) \simeq 1.2\%$~ or~
${\cal B}(B^0 \to D^*(2010)^+\; {\bar D}^{*}(2007)^0 K^-) \simeq 1.2\%$~
are not very small, while the available kinetic energy remains below $1\;$GeV.

Fortunately, we have already exhausted the case with no hard gluons in
Sec.~\ref{subsec:ccsoft}. Thus, the non-perturbative contributions that we are
now after must be suppressed by at least one power of $\alpha_s$. Such a
suppression of the annihilation channels is not surprising because the
considered states can decay to open charm without any hard gluon
emission/exchange. In fact, when these states are really ``long-distance''
ones, $\alpha_s$ is not the only dumping factor. The $c$ and $\bar c$
wavefunctions are then diluted over distances of order at least
$\Lambda^{-1}$, so the annihilation probability must be correspondingly
suppressed. By analogy to the $B$-meson decay constant that scales like
$(\Lambda/m_b)^{3/2}$ due to the meson size, we may conclude that $c\bar c$
annihilation probability for the ``long-distance'' states is down by
$(\Lambda/M)^{3/2}$, where $M$ is any of the hard scales in the problem ($2
m_c$, $2 E_\gamma$, $m_b$). To be conservative, let us skip the power ``3/2''
and assume for error estimates that the considered unknown non-perturbative
corrections to ${\cal B}_{s\gamma}$ are of order ${\cal O}(\alpha_s
\Lambda/M)$.

An uncertainty due to such corrections at the 5\% level has been assumed in
the SM prediction in Eq.~(\ref{sm}), just because $\alpha_s(m_b) \simeq 0.2$,
and $\Lambda/M$ is not much smaller. This error has been added in quadrature
to the other ones, so it should be interpreted as a ``theoretical 1$\sigma$''
rather than a strict upper bound on the size of such effects. Actually, the
uncertainty might have been overestimated because, as shown in
Fig.~\ref{fig:rings}, the ${\cal O}(\alpha_s \Lambda/M)$ correction
corresponds to the $b \to s g\gamma$ channel that poorly interferes with the
leading term; the corresponding perturbative interference has no peak near
$E_\gamma^{\rm max} \simeq m_b/2$.  On the other hand, the hard $c\bar c s \to
s\gamma$ subprocess that might lead to effective interference is suppressed by
an additional power of $\Lambda/M$ due to wave-function
dilution. Nevertheless, any more optimistic error estimate would need to be
based on a detailed analysis that seems to be rather difficult.

\section{Final remarks}

In this short status summary, the main stress has been put on the rarely
discussed issues which, however, are responsible for the main theory
uncertainty in ${\cal B}_{s\gamma}$. A far as the perturbative calculations
are concerned, the reports in Ref.~\cite{Misiak:2008ss} remain still
up-to-date. Several recent analyses of the photon energy spectrum near
endpoint can be found in Ref.~\cite{Ligeti:2008ac}. They are essential for the
$E_0$ extrapolation that has been mentioned in Sec.~\ref{sec:intro}.

\section{Acknowledgements}

I would like to thank the Ustro\'n meeting organizers for invitation and
hospitality. This work has been supported in part by the Polish Ministry of
Science and Higher Education as a research project N~N202~006334 (in years
2008-11), by the EU-RTN Programme ``FLAVIAnet'' (MRTN-CT-2006-035482), and by
the EU FP6 Marie Curie Research \& Training Network ``UniverseNet''
(MRTN-CT-2006-035863).

\end{document}